# Creating music based on quantitative data from variable stars

Droppelmann, C.A.[1] and Mennickent, R.E.[2]

[1]Robarts Research Institute, Western University, London, ON, Canada. Email: cdroppel@uwo.ca

[2]Astronomy Department, Universidad de Concepción, Casilla 160-C, Concepción, Chile. Email: rmennick@udec.cl

**Abstract.** In this work we show a technique that allows for the musical interpretation of the brightness variations of stars. This method allows composers a lot of freedom to incorporate their own ideas into the score, based on the melodic line generated from the quantitative data obtained from the stars. There are a wide number of possible applications for this technique, including avant-garde music creation, teaching and promotion of the association between music and science.

Musical adaptation of the Universe can be understood as a scientific and artistic adventure. When we convert the changes of brightness of variable stars into music, two disciplines converge – astronomy allows us to detect, to record and to interpret the changes in stars, and music allows us to represent those light fluctuations through art. This last step, entirely artistic, requires the creation of a method that allows us to analyze and interpret the energy flows received from stars using musical parameters, including pitch and rhythm. This activity requires intelligence and a good sense of aesthetics.

Few attempts of sound creation from stellar light curves have been made until now. Some examples of sonification – from the Kepler Special Mission – derived light curves from solar-type stars, red giants, cataclysmic variables, and eclipsing binary stars[1]. These sonifications have received the attention of the New York Times and were featured in an article entitled “Listening to the Stars”[2]. Additionally, there are few groups working in the sonification of diverse astronomical events, including gamma ray bursts (for a list of examples see Table 1). However, in the context of stellar light curves sonifications, they do not necessarily represent an artistic representation of the light curves, nor are they musically attractive. In this article, we propose a new method of converting star light curves into music, so that the resulting music would be more aesthetically pleasing to the human ear. This method is a new way of creating music and has the potential to become a powerful tool for the development of avant-garde music, for musical teaching, for providing a listening experience to visually impaired people, based on astronomical data and phenomena, and for encouraging public interest in astronomy.

**Table 1**. Examples of astronomical data sonification.

| **Astronomical data sonification** | **Links** |
|---|---|
| Exoplanets (discrete sounds and multiple instruments). Data sonification using Python MIDI-based code SONIFY. (Erin Braswell) | https://osf.io/vgaxh<br>http://astrosom.com/Aug2018.php |
| Gamma ray bursts (discrete sounds and multiple instruments). (Sylvia Zhu) | https://blogs.nasa.gov/GLAST/2012/06/21/post_1340301006610/ |
| Gamma ray bursts. Sonification based on ALMA astronomical spectra. (Tanmoy Laskar) | https://public.nrao.edu/the-sound-of-one-star-crashing-haunting-melody-from-the-death-of-a-star/ |
| Flaring Blazar (Several techniques). (Matt Russo and Andrew Santaguida) | http://www.system-sounds.com/the-creators/<br>https://svs.gsfc.nasa.gov/12994 |

To create our first composition, we choose the star RV Tauri ($\alpha_{2000}$ = 04:47:06.7, $\delta_{2000}$ = +26:10:45.6), the prototype of the variable stars dubbed RV Tau stars, characterized by almost regular bright oscillations. These oscillations are characterized by alternating brightness minimums that are modulated on irregular fashion. The AAVSO data available on-line (www.aavso.org) indicates that RV Tauri has a period of around 78 days and shows two maxima at V magnitude around 9.0, a minimum around magnitude 10.0, and another minimum about 0.5 magnitudes fainter. This behavior cannot be fully explained, but is believed to be caused by chaotic stellar pulsations. Another reason we chose this star is because we had access to high-quality bright variation data that was collected over several years. The light curve was obtained from The All Sky Automated Survey catalog[3]. The curve consists of 180 photometric magnitudes in the V band, qualified as A-type (of the highest quality) obtained between Heliocentric Julian days 2452621.66922 and 2455162.70827, scilicet 7 years. The data is presented in two columns, Heliocentric Julian day versus V magnitude. The curve, presented in the Figure 1, shows the light change of the star versus time.

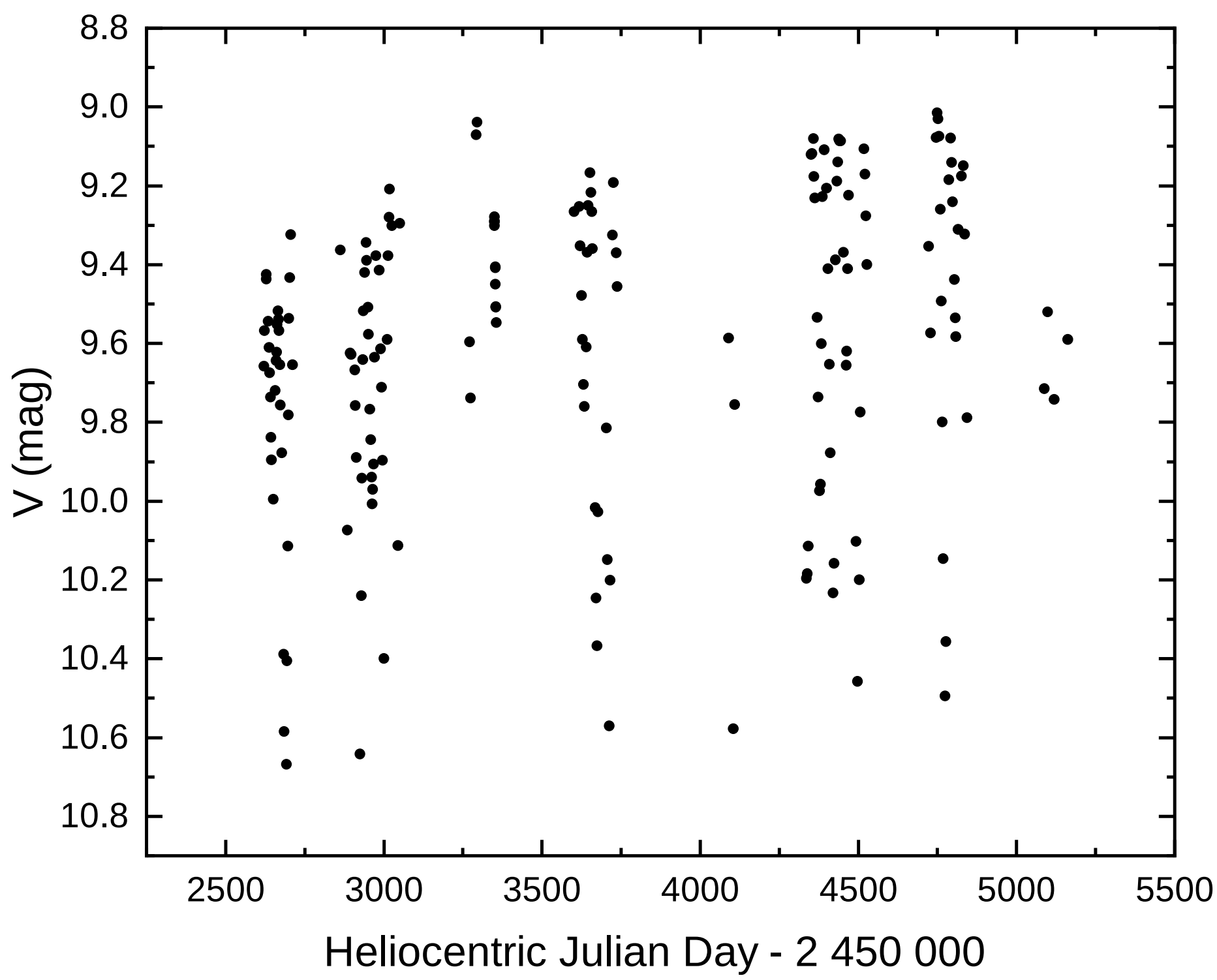

**Figure 1.** Light curve of RV Tauri.

To create the musical composition, the magnitudes of the star were converted into musical notes using the following equation:

$$M_N = \frac{(M - M_{min})}{(M_{max} - M_{min})} \times S$$

Where: $M_N$= Normalized magnitude

$M_{min}$ = minimal magnitude

$M_{max}$ = maximal magnitude

$M$ = magnitude in the chosen Heliocentric Julian day

$S$= Number of semitones. In this case 24 semitones (2 octaves)

As a consequence, one unit of normalized magnitude corresponds to one semitone. We assigned only 24 semitones to avoid excessive chromaticism; however, this parameter could be changed by a composer to include more chromaticism or include microtonality, if desired. Also, the composer could choose between using high magnitude values to higher tones or vice versa only by assigning the higher normalized value to the higher note of the 24 semitones or the lower normalized value to the higher tone (see Table 2). By artistic reasons in this work we used higher normalized values to higher tones (Option 1 in Table 2).

**Table 2.** Assignment of tones after normalization of magnitude values using a 24 semitones chromatic scale.

| **Tone** (chromatic scale) | $M_N$ **Option 1** (higher note = higher mag. value) | | $M_N$ **Option 2** (higher note = lower mag. value) | |
|---|---|---|---|---|
| C | 0 to 0.99 | (octave higher) | 24 | (octave higher) |
| C♯/D♭ | 1 to 1.99 | 13 to 13.99 | 23 to 23.99 | 11 to 11.99 |
| D | 2 to 2.99 | 14 to 14.99 | 22 to 22.99 | 10 to 10.99 |
| D♯/E ♭ | 3 to 3.99 | 15 to 15.99 | 21 to 21.99 | 9 to 9.99 |
| E | 4 to 4.99 | 16 to 16.99 | 20 to 20.99 | 8 to 8.99 |
| F | 5 to 5.99 | 17 to 17.99 | 19 to 19.99 | 7 to 7.99 |
| F♯/G♭ | 6 to 6.99 | 18 to 18.99 | 18 to 18.99 | 6 to 6.99 |
| G | 7 to 7.99 | 19 to 19.99 | 17 to 17.99 | 5 to 5.99 |
| G♯/ A♭ | 8 to 8.99 | 20 to 20.99 | 16 to 16.99 | 4 to 4.99 |
| A | 9 to 9.99 | 21 to 21.99 | 15 to 15.99 | 3 to 3.99 |
| A♯/B♭ | 10 to 10.99 | 22 to 22.99 | 14 to 14.99 | 2 to 2.99 |
| B | 11 to 11.99 | 23 to 23.99 | 13 to 13.99 | 1 to 1.99 |
| C | 12 to 12.99 | 24 | 12 to 12.99 | 0 to 0.99 |

A preliminary tone assignment for the RV Tauri data was created following the information of Table 1. Then, after the definition of the rhythm for the music, it was assigned a key with 3 sharp alterations (A major), that fitted the chromatism of the melodic line generated by the stellar information, while keeping the pitches of the notes. Then, considering artistic reasons (to suit the capabilities and range of a string orchestra) the melodic line was transposed to E major (4 sharps) and finally, accidentals were eliminated enharmonically to simplify the interpretation by musicians.

Consequently, the range of the stellar melodic line in the final score goes from G4 to G6.

The rhythm was assigned based on the time interval between magnitude measurements in Heliocentric Julian days using the following equation:

$$t_N = t_n - t_{n-1}$$

Where: $t_N$= Normalized time interval

$t_n$ = time of selected magnitude measurement in Heliocentric Julian days

$t_{n-1}$ = time of previous magnitude measurement in Heliocentric Julian days

Note: The first Heliocentric Julian day has assigned a normalized value of 1.

Once obtained the normalized time interval, we assigned the rhythm based in a range that goes from eighth notes to whole notes (see Table 3). This method allows for the creation of a rich and diverse rhythm pattern for the score, including the incorporation of intermediate-length musical notes (dotted notes) under the composer criteria when normalized time interval values are close to the next assignation interval. Any gap in the brightness measurement created longer notes. However, the creation of rests is under criteria of the composer. In the music conceived for this paper no rests were used since our goal was to produce a feeling of continuity in the music. This criterion was used because a variable star only has changes in its brightness and not interruptions.

**Table 3.** Assignment of rhythm after normalization of time intervals between Heliocentric Julian days for RV Tauri data.

| Normalized time interval | Note assigned |
|---|---|
| 0 to 0.99 | Eighth note |
| 1 to 3.99 | Quarter note |
| 4 to 15.99 | Half note |
| 16 to 63.99 | Whole note |
| Over 64 | Ligated whole notes |

Note: If more rhythmic richness is desired, the assignment can be initiated from thirty-second or sixteenth note. Also, a specific interval for dotted notes could be assigned.

Once the melodic line from the stellar data was created (Figure 2), the orchestral arrangement started with a pedal note based on the tonic note of the key use for the music (E major), and was structured as a four-voice canon with a major sixth interval. It should be noted that at this stage, the composer has complete freedom to arrange the composition based on the stellar melodic line. The score present here is only one of infinite possibilities and denotes the richness and potential of this method for the creation of new music, where even one star allows for the creation of an immense range of new music.

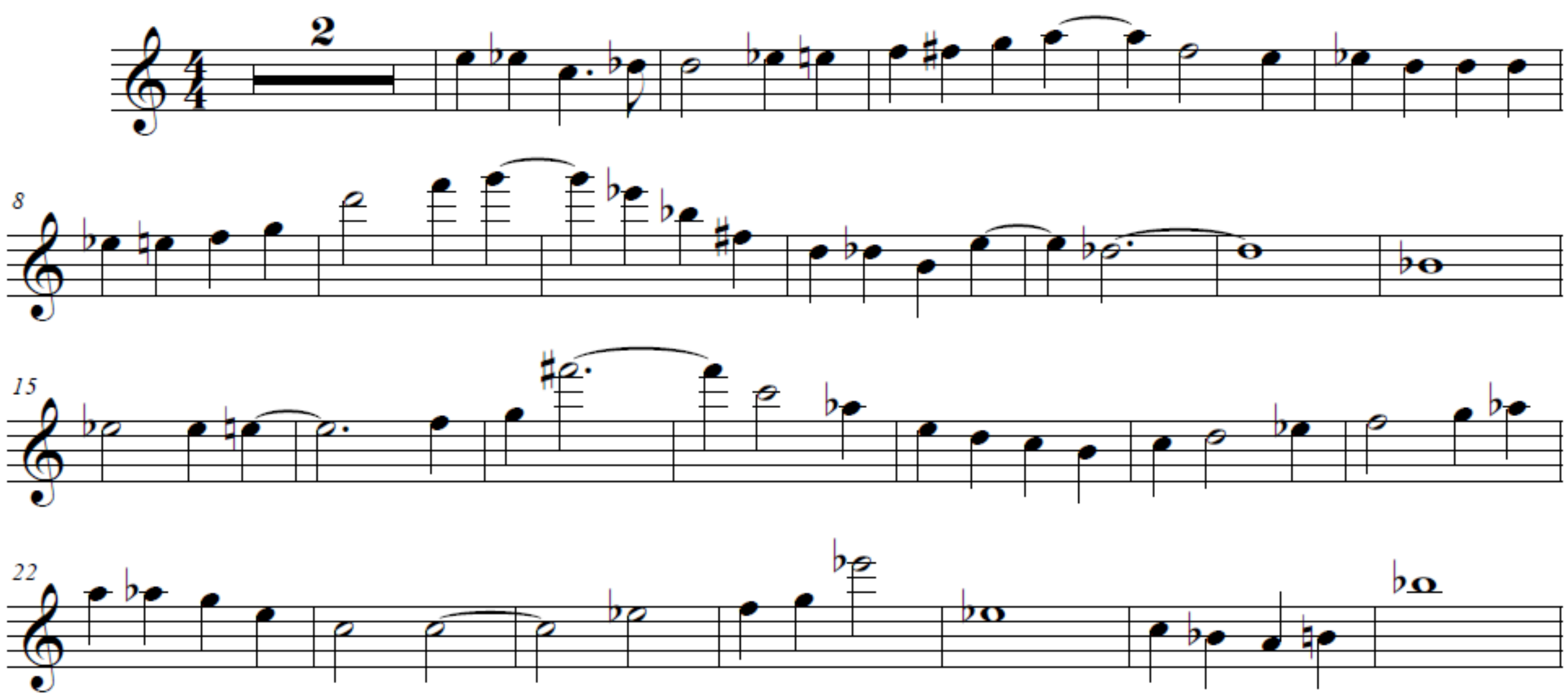

**Figure 2.** First 28 measures of the melodic line generated for RV Tauri.

The score was made for a string orchestra, including first and second violins, violas, cellos and double basses. The tempo was set at 50 bpm, and used simple time signature (4/4), equal temperament and baroque pitch (A= 415 Hz). Finale 25 software and the Garritan library were used to create the score and the audio file. The music starts with the pedal note being played by double basses. Then, the melodic line (the music derived from the star light curve), is introduced by the first violins in the third measure. In the fourth measure, the second violins begin the canon as the second voice. Then, the violas begin in the fifth measure and finally, the cellos in the sixth measure. From there, the musical body is developed by following melodic line derived from the star, until the last musical note corresponds to the last magnitude measured. At this point, the canon is ended to give a sense of completion, as is necessary for a musical score (Figure 3). The full score and the audio file can be downloaded as supplementary files.

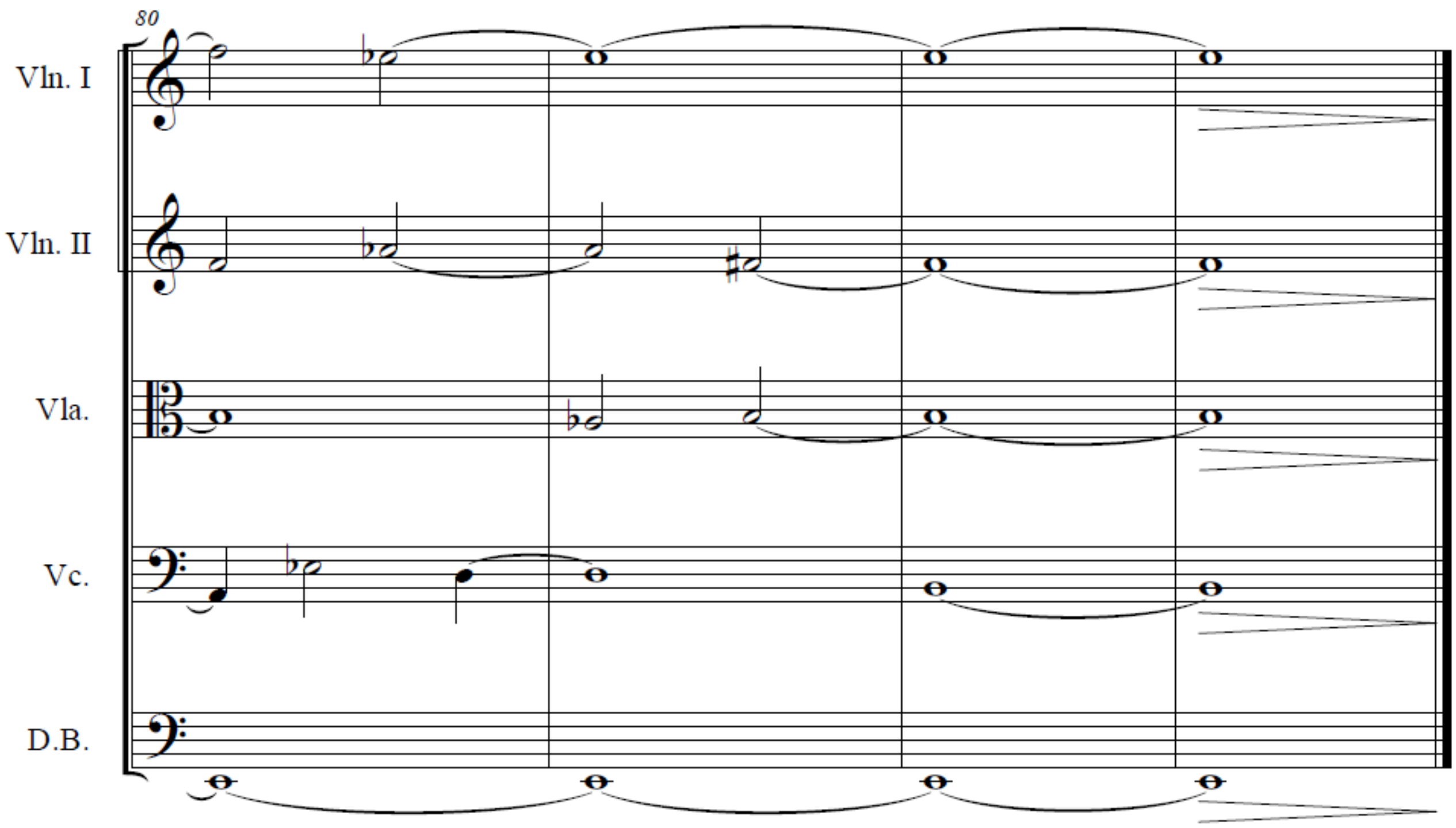

**Figure 3.** Last 4 measures of the orchestral score based in the melodic line generated for RV Tauri.

## Conclusions

Here, we have developed a technique that allows for the musical interpretation of the brightness variations of stars. The advantage of this method is that it allows composers a lot of freedom to incorporate their own ideas into a score, based on the melodic line obtained from stars. There are a wide number of possible applications for this technique, including avant-garde music creation, teaching and promotion of the association between music and science. It could even be attractive for persons with visual impairments, who are interested in astronomy, to 'hear' the universe rather than 'see' it.

## References

1 http://kepler.nasa.gov/multimedia/Audio/sonification/

2 http://www.nytimes.com/2011/01/31/science/space/31star.html?_r=2

3 Pojmanski, G. "The All Sky Automated Survey". *Acta Astronomica*, 1997, 47: 467

## Supplementary information:

1) Link to music (for listening or download):

https://my.pcloud.com/publink/show?code=XZDua17ZafxBcBqhi2HOQgi0kVlR0uh6TPky

2) Full score of RV Tauri for string orchestra (attached)

Score

# RV Tauri

Music: Cristian Droppelmann
Astronomic data: Ronald Mennickent

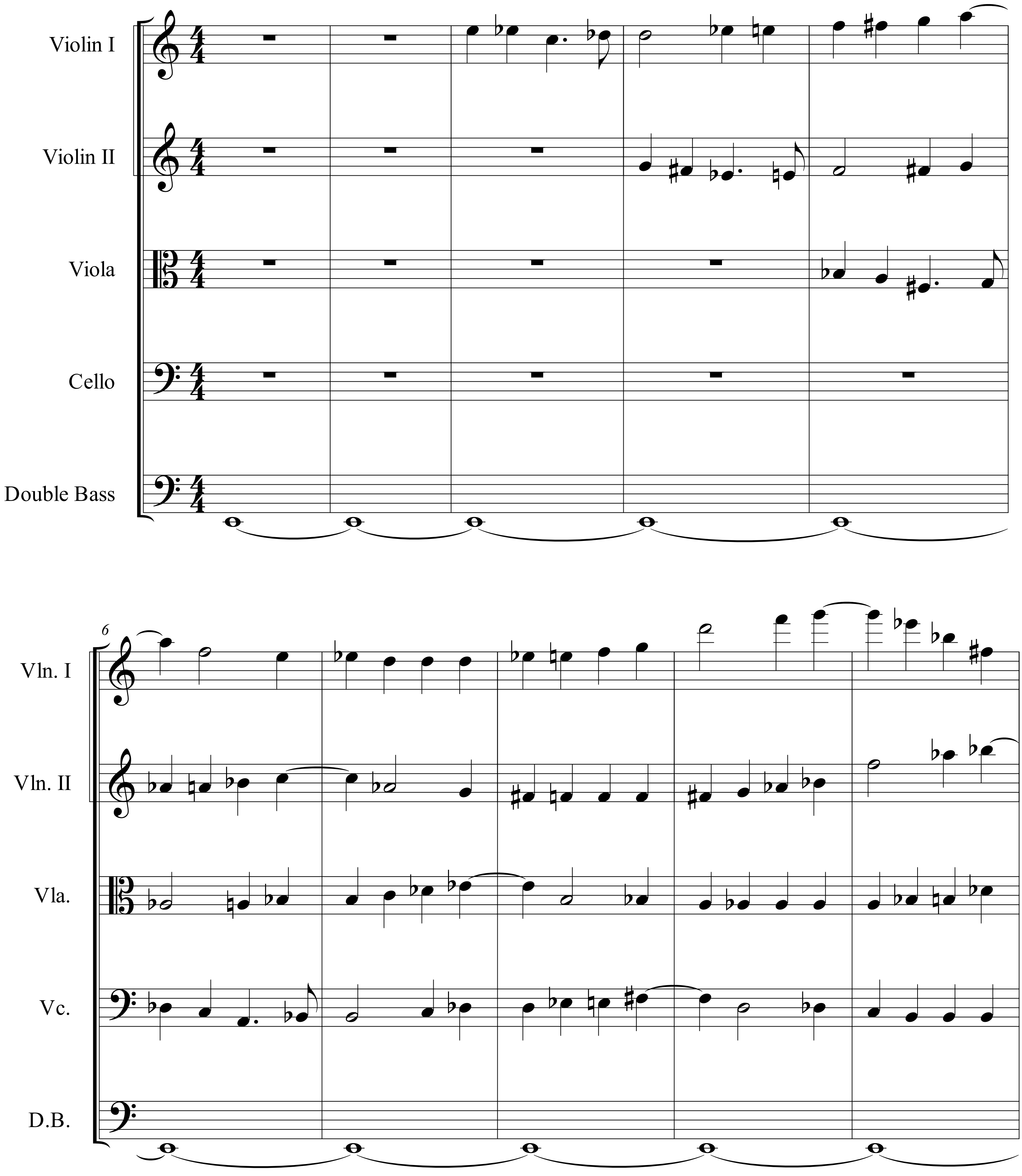

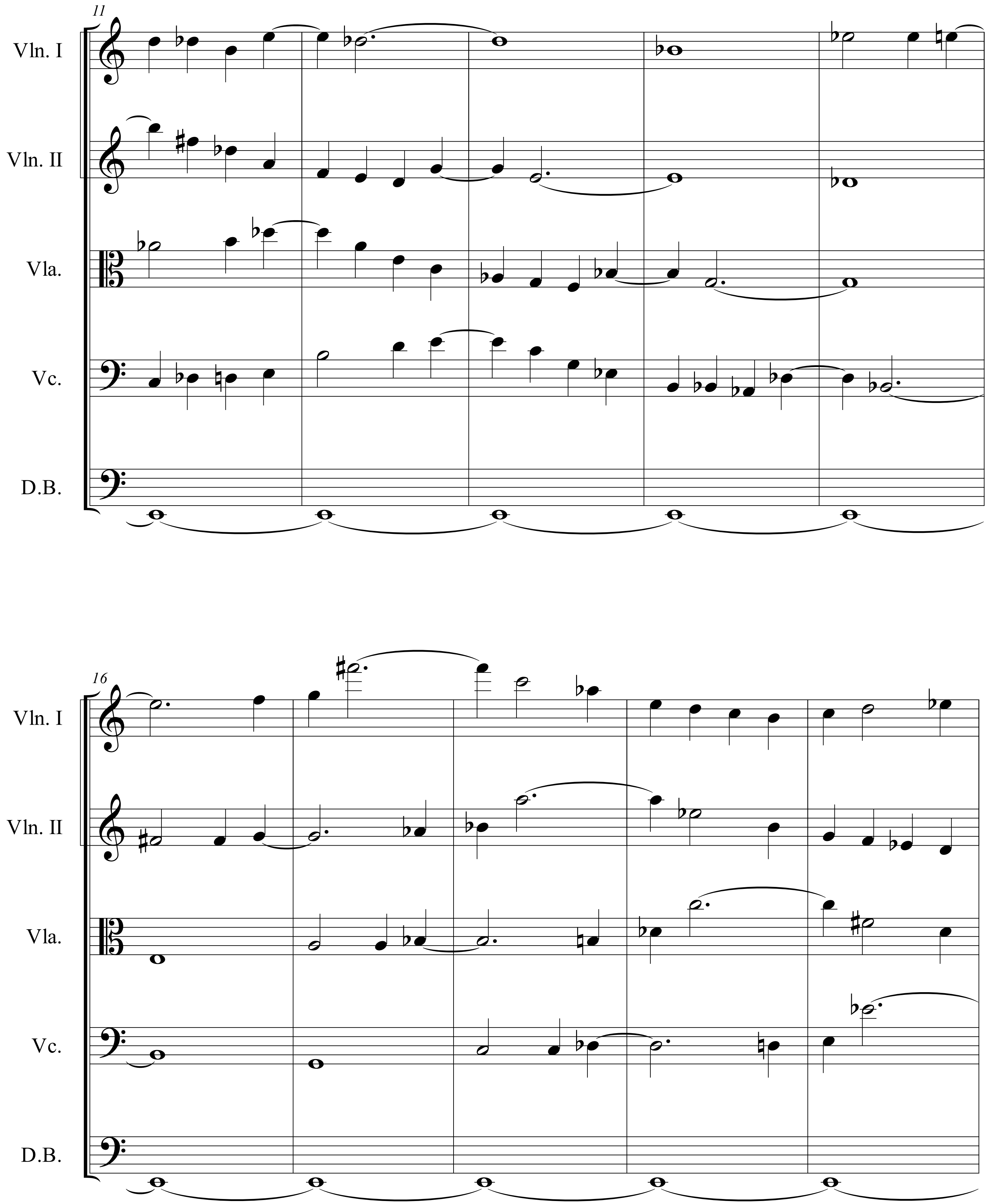
11
Vln. I
Vln. II
Vla.
Vc.
D.B.
16
Vln. I
Vln. II
Vla.
Vc.
D.B.

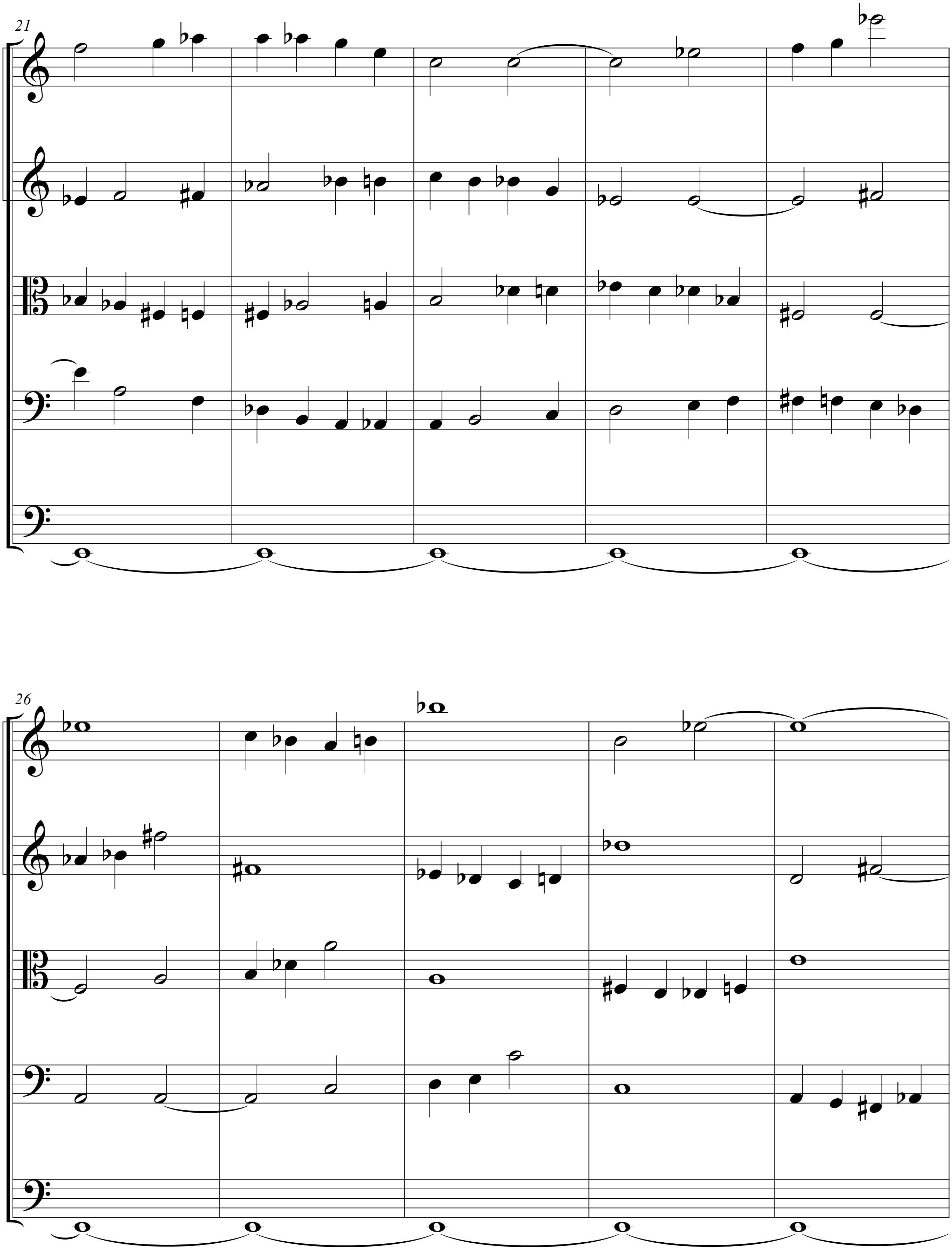
21
Vln. I
Vln. II
Vla.
Vc.
D.B.
26
Vln. I
Vln. II
Vla.
Vc.
D.B.

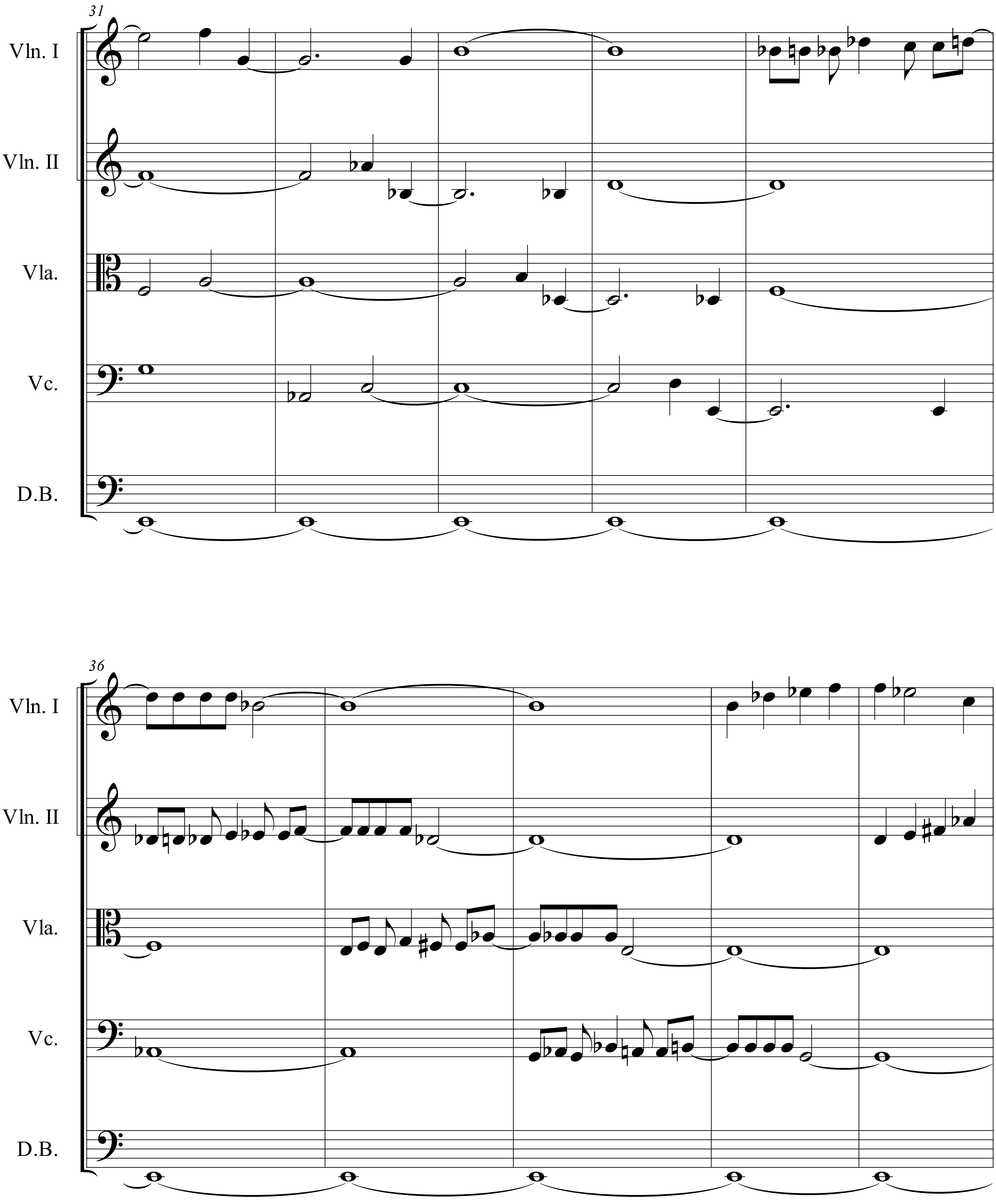
31
Vln. I
Vln. II
Vla.
Vc.
D.B.
36
Vln. I
Vln. II
Vla.
Vc.
D.B.

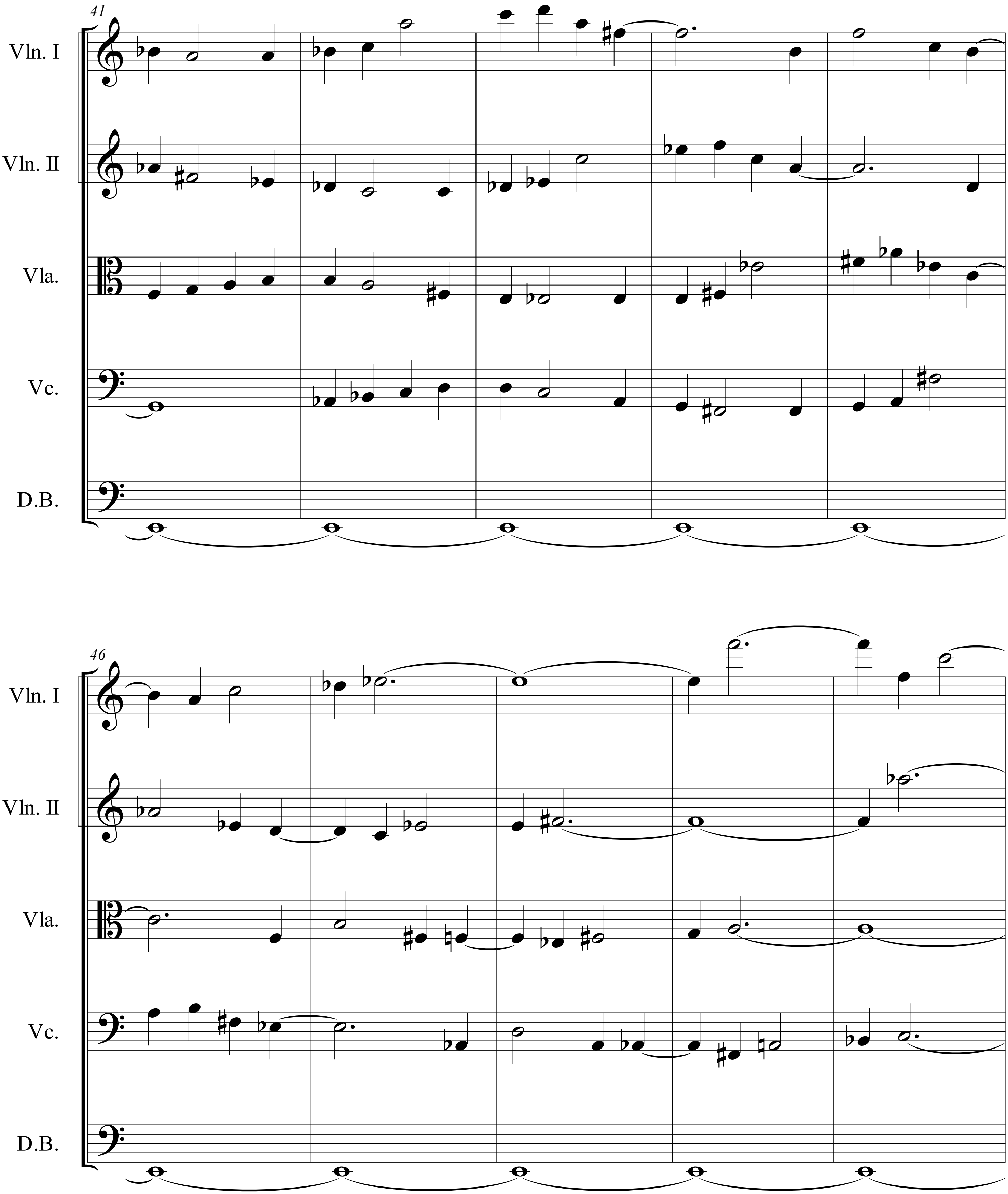
41
Vln. I
Vln. II
Vla.
Vc.
D.B.
46
Vln. I
Vln. II
Vla.
Vc.
D.B.

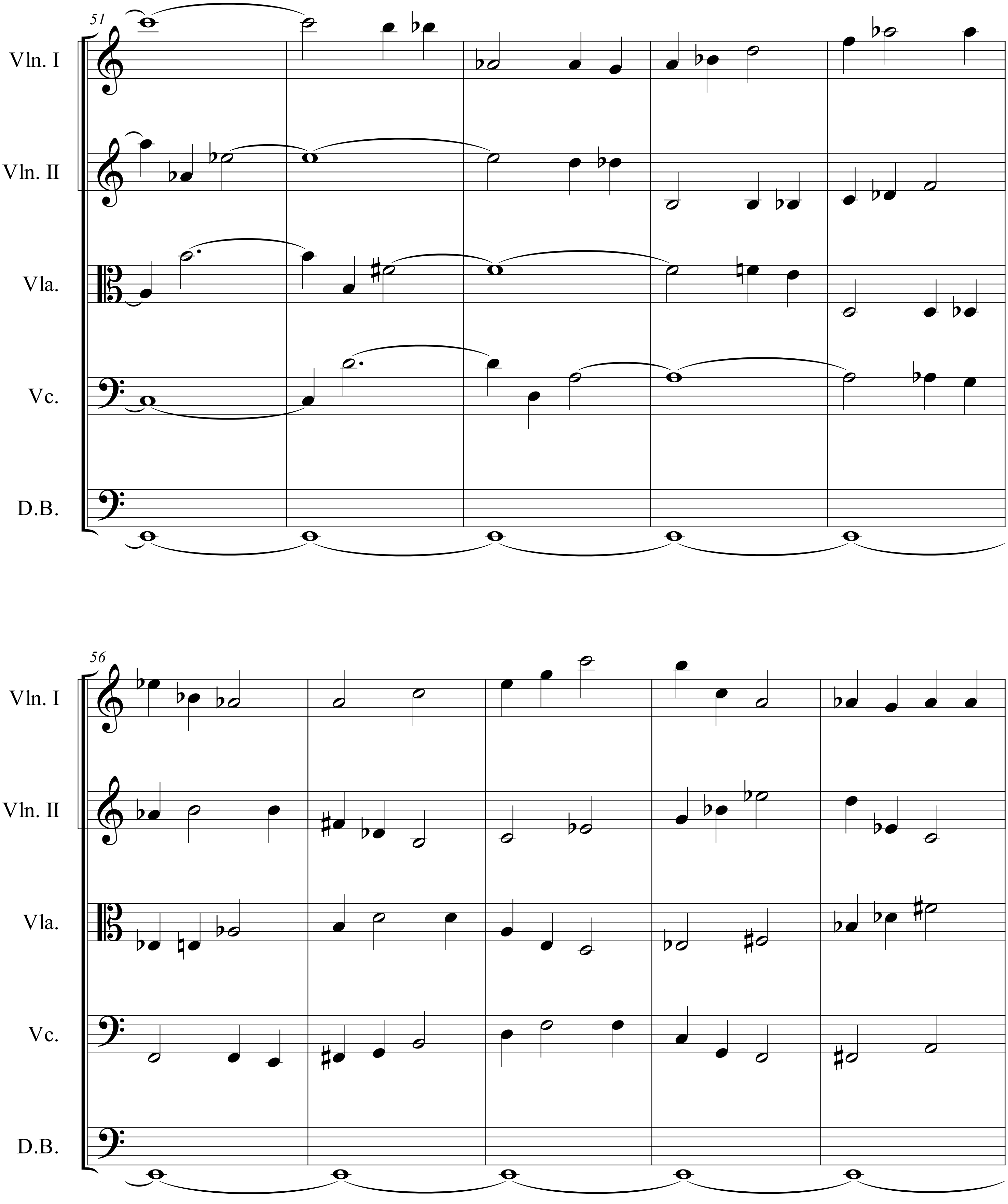
51
Vln. I
Vln. II
Vla.
Vc.
D.B.
56
Vln. I
Vln. II
Vla.
Vc.
D.B.

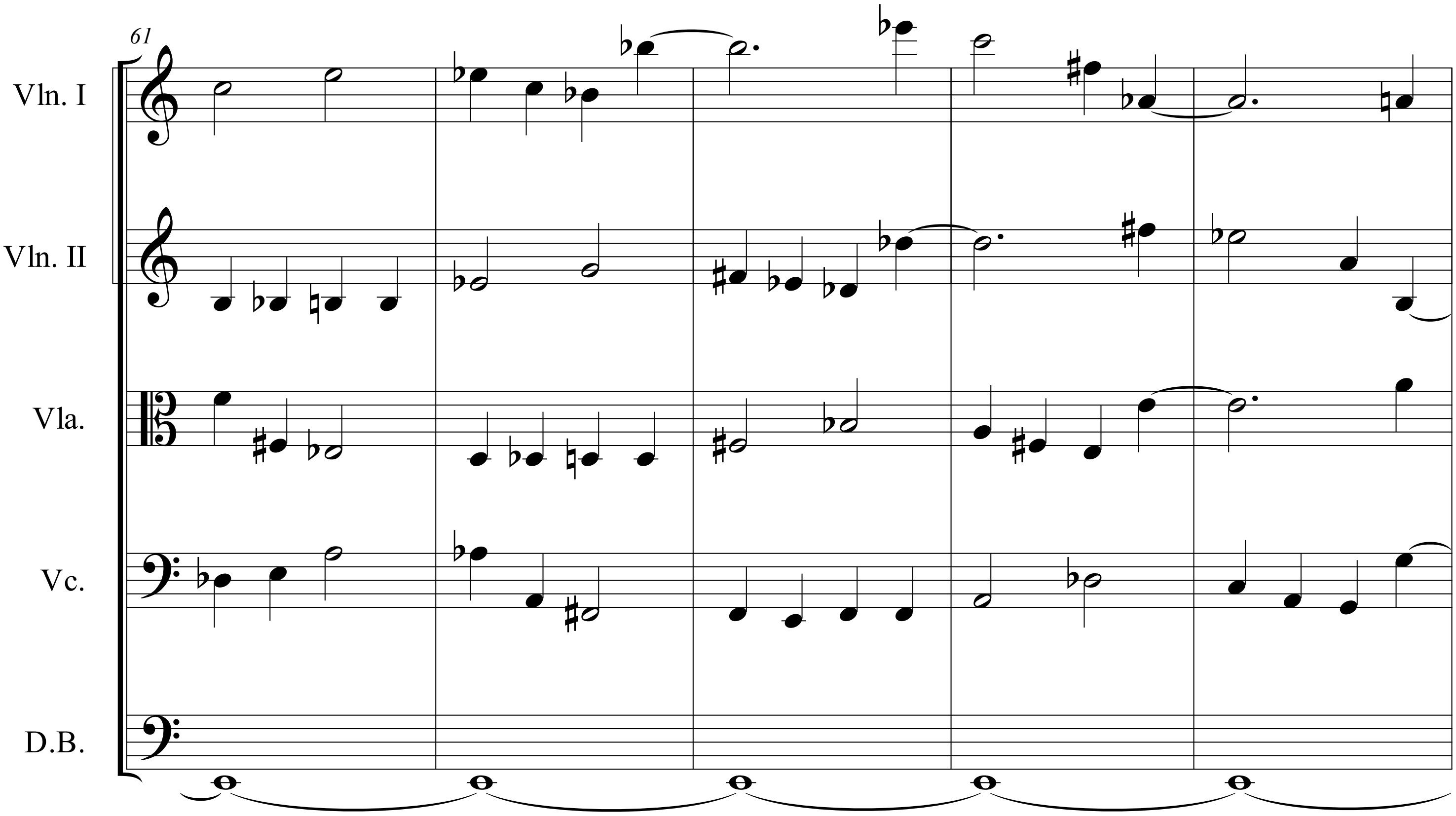
61
Vln. I
Vln. II
Vla.
Vc.
D.B.

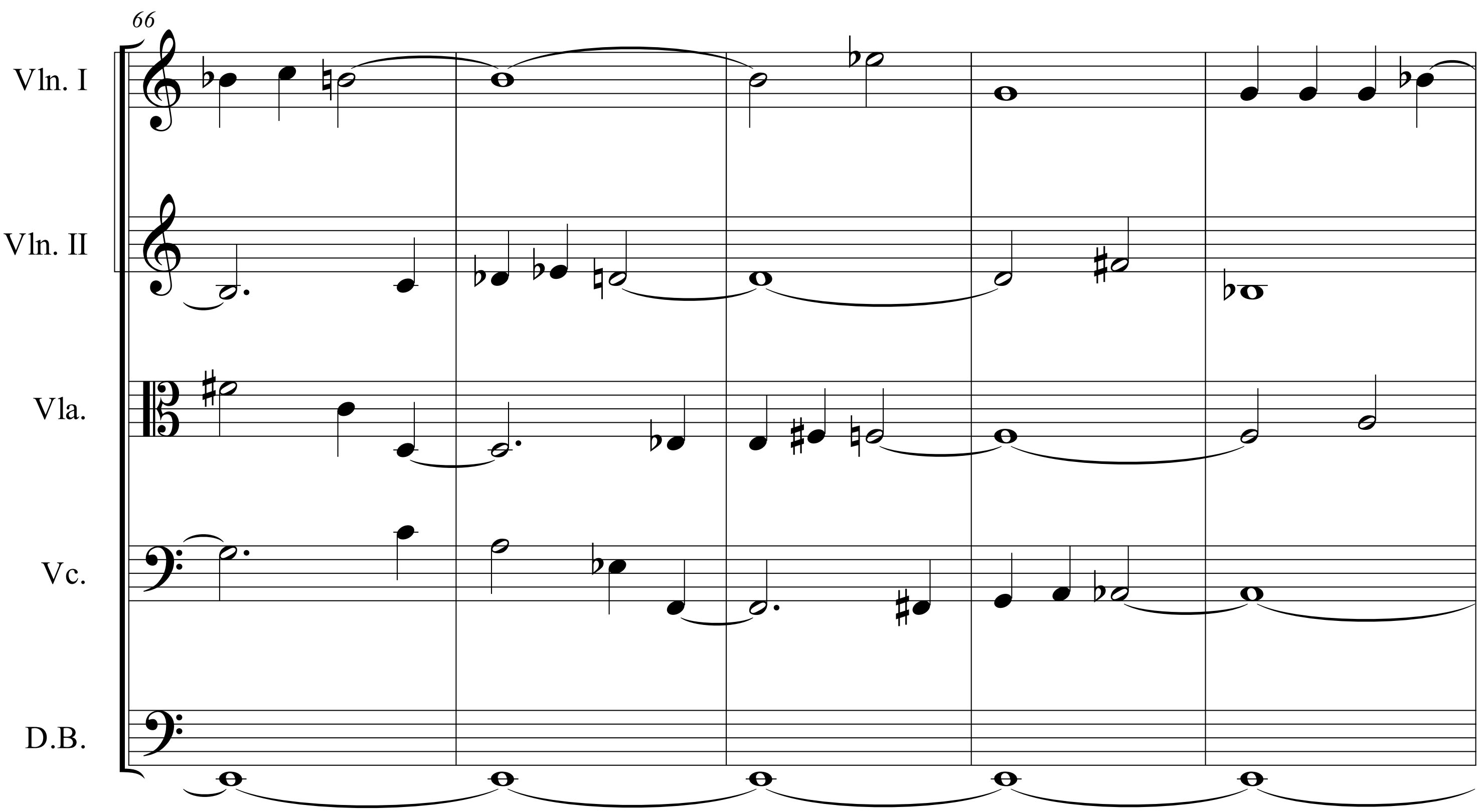
66
Vln. I
Vln. II
Vla.
Vc.
D.B.

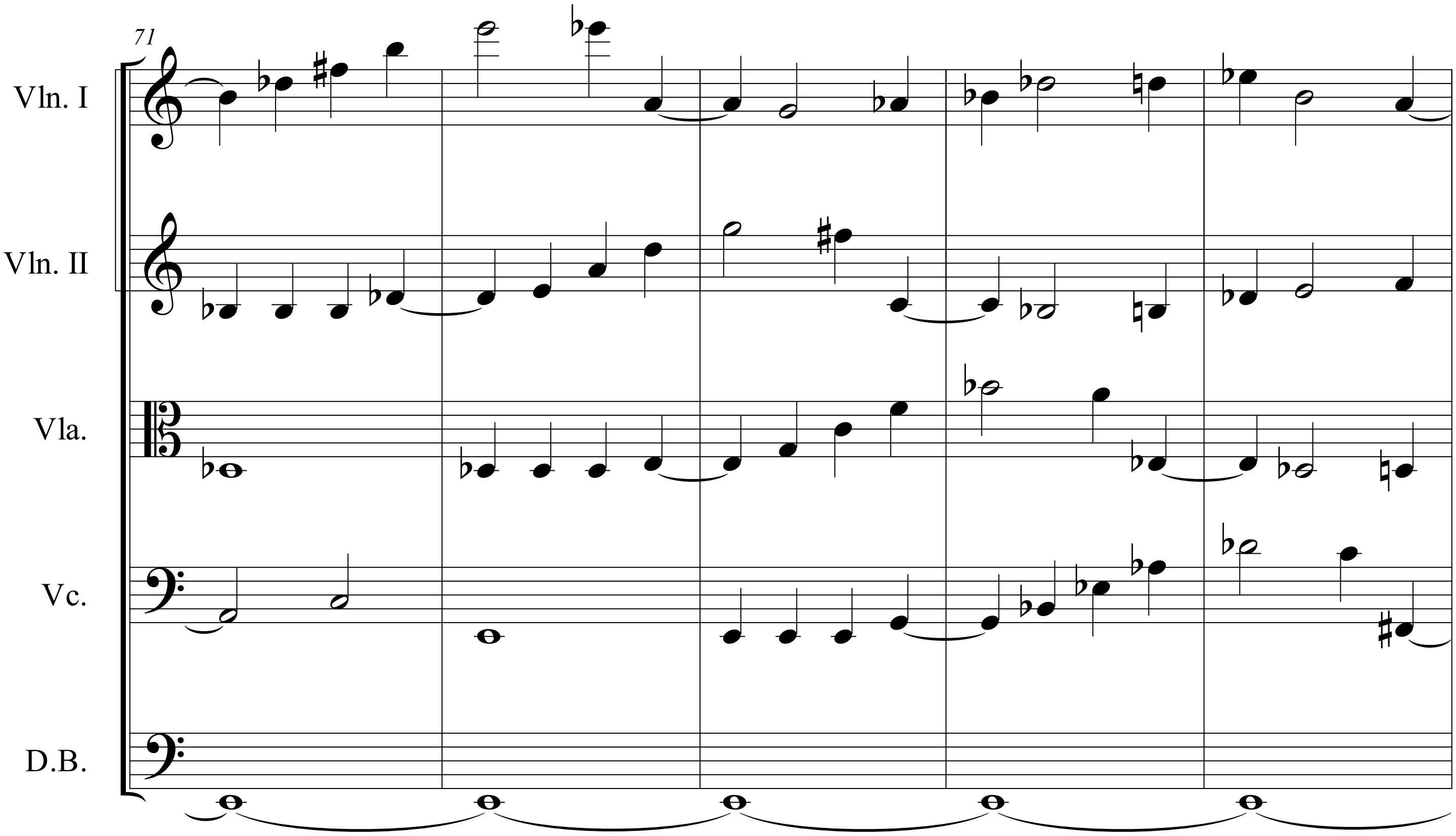
71
Vln. I
Vln. II
Vla.
Vc.
D.B.

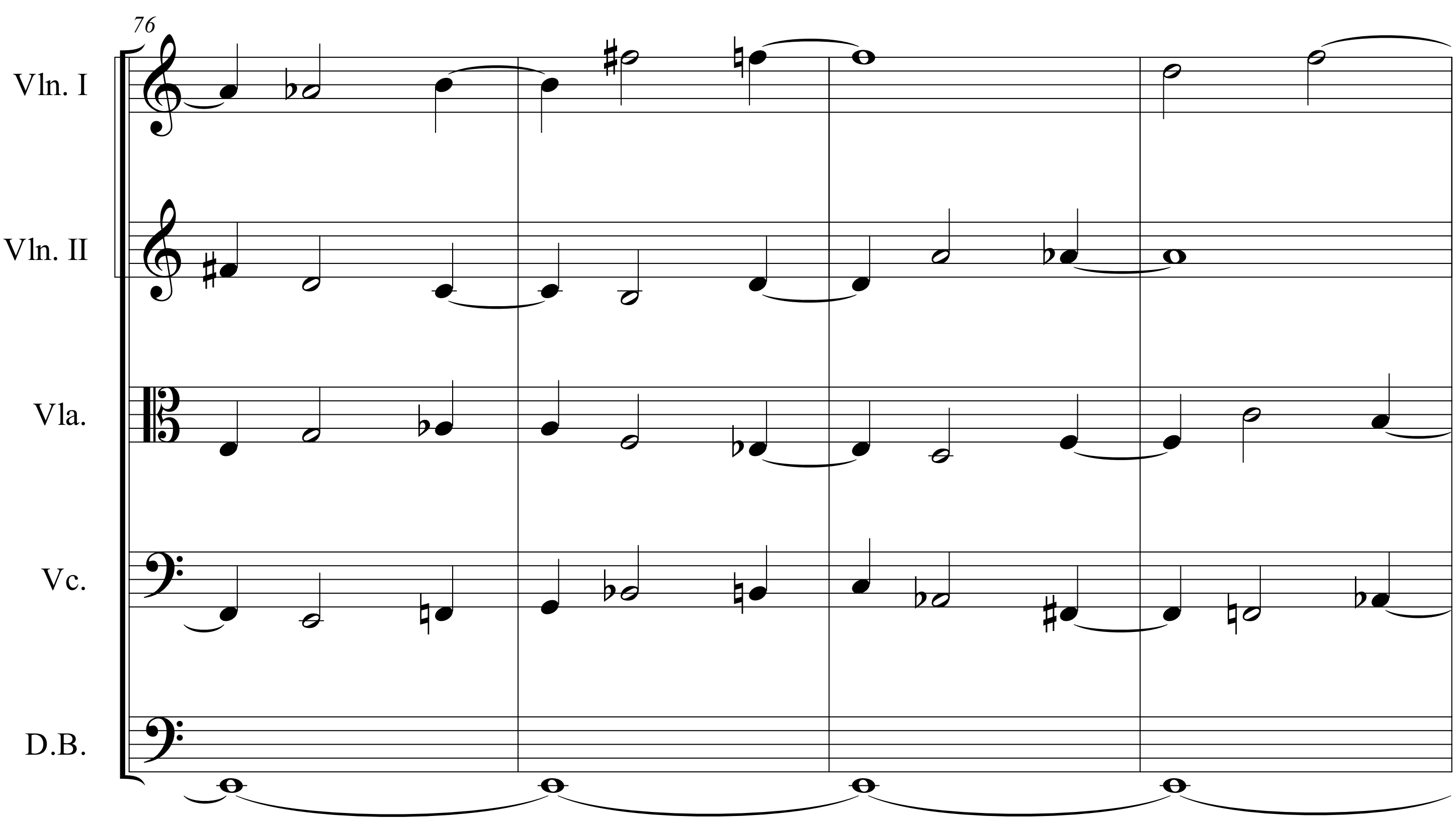
76
Vln. I
Vln. II
Vla.
Vc.
D.B.

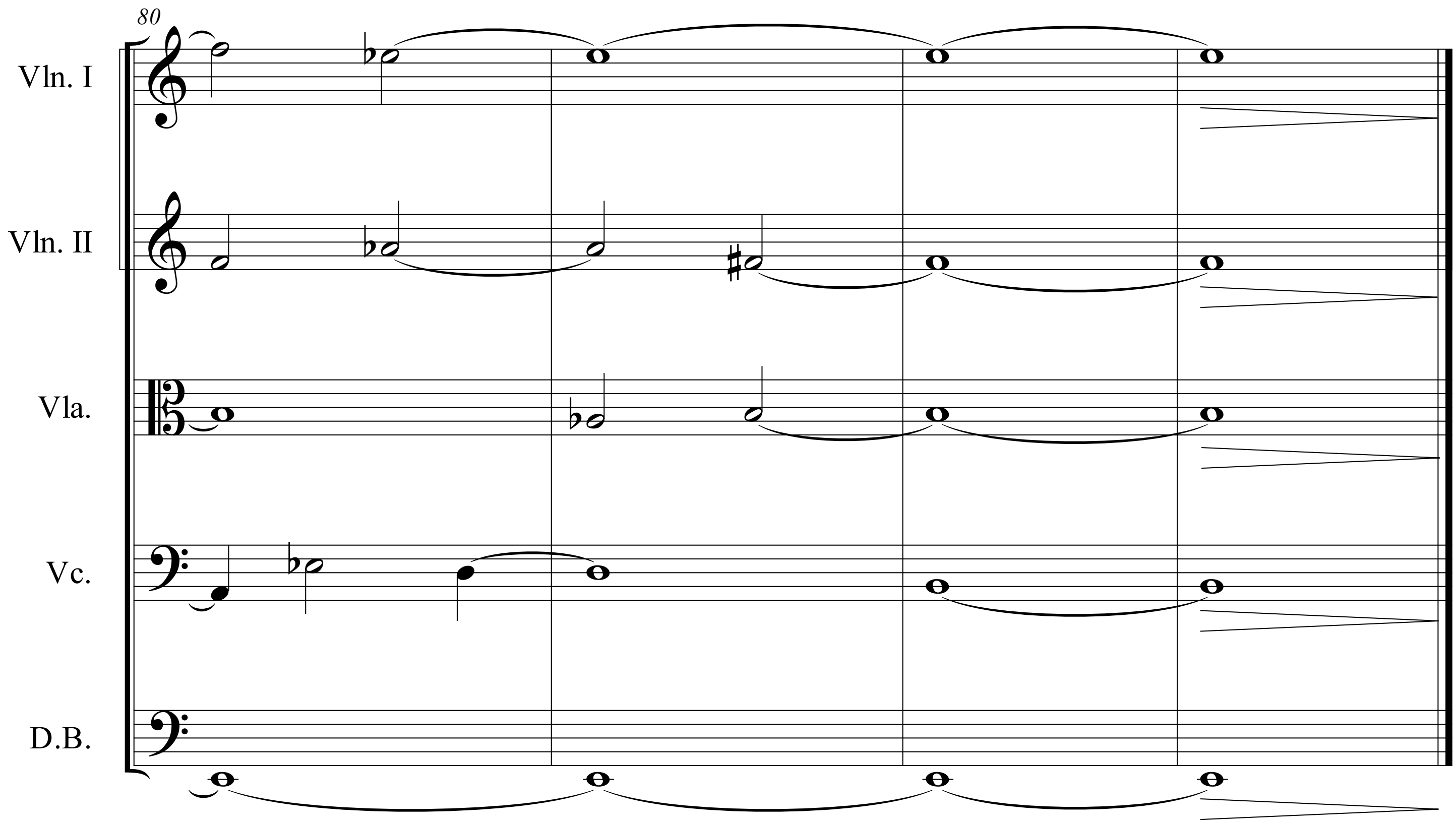
80
Vln. I
Vln. II
Vla.
Vc.
D.B.